\documentstyle[epsf,aps,prl]{revtex}

\begin{document}   
\draft
     
\wideabs{

\title{Analogy between two-dimensional dislocation
systems and  layered superconductors } 
 
\author{J. Cserti}  
\address{E\"otv\"os University, Department of Physics of Complex Systems, 
H-1088 Budapest, M\'uzeum krt.  6-8, Hungary}   

\date{\today}   

\maketitle 
\begin{abstract}  We propose a consistent treatment of the applied external
stress on a two-dimensional dislocation systems by using the analogy
between our model and the layered superconductors. Using the results of 
a mathematically rigorous, real-space 
renormalization-group (RG) calculations in the latter model we extend the
original model developed by Khantha {\it et al.} 
From our recursion relations a nonlinear equation is derived 
for the transition temperature as a function of the 
applied stress. Our results are compared with the predictions of Khantha's
model for three particular materials and the reasons for the
deviations are discussed. Possible extensions of our model are outlined.
\end{abstract}

\pacs{PACS numbers: 05.20.-y, 61.72.Lk, 62.20.-x, 64.70.-p}

}



In the context of dislocation pattern formation  the collective behavior 
of the dislocation systems seems to play a fundamental role and a 
new approach has been used to determine the distribution function of 
the dislocation systems \cite{Groma}. 
However, in this model the temperature dependence has not been included.  
Recently, a new type of mechanism of the cooperative generation  of the 
dislocations has been proposed by Khantha, Pope and Vitek (hereafter 
referred as to KPV model) \cite{KhPV-PRL}-\cite{KhPV-Engin}. 
The KPV model of the unstable generation of dislocations driven by 
thermal fluctuation and aided by an applied stress on the solid is 
based on the theory of the screening of topological defects (vortices) 
developed by Kosterlitz and Thouless \cite{KT,Kosterlitz} 
and Berezinskii \cite{Berez} (hereafter referred as to KTB). 
The KTB type transition of the topological defects (vortices) 
has already been well established in several physical systems such as 
dislocation mediated melting in 2D crystal 
\cite{NelsonHalperin,Young}, layered high-temperature 
superconductors \cite{Pierson,Pierson-J}, Coulomb gas \cite{Minnhag}, 
superfluid ${}^4$Helium films, 
2D X-Y model, liquid crystals (for review see \cite{Lubensky}).
The KTB phase transition seems to play a fundamental role in the behavior 
of some mesoscopic devices like double layer quantum Hall 
systems \cite{Mats}, where the statistical mechanics of the so called merons 
analogous to that of the vortices.

In the KTB type transition the bounded dislocation pairs unbind (free 
dislocations appear in thermal equilibrium) at a certain temperature where 
an universal jump in shear modulus takes place. As an extension of the 
KTB theory in the KPV model the effect of applied stress is included 
additionally through a mean-field type calculation. In the model one 
of the main results is the recursion equations for the coupling constant 
and the fugacity which relate to the strength of interaction between 
dislocations and the average number of dislocations in the system, 
respectively. Due to the applied external stress the change in fugacity 
was  taken into account in the derivation of the recursion relations 
and it results in a modified recursion equation for 
the fugacity in comparison with that of the KTB theory. 

However, the stress not only increases the fugacity (and consequently the 
density of dislocation pairs) but indirectly affects the coupling constant 
which can be understood as follows. The number of
dislocation pairs increases with increasing stress. 
Larger number of dislocation pairs 
makes more effective the screening of the interaction between dislocations  
resulting in a decrease in the coupling constant. It is clear that in a 
{\it consistent} treatment of the effect of the stress one 
needs to take into account this {\it indirect effect} of the stress 
which was omitted in the KPV model.

The vortex gas in the layered superconductors shows many common
aspects with the above discussed dislocation systems in the sense that 
the character of the interaction between  vortices and  dislocations
can be identical in some cases. The Hamiltonian used in the KPV model is
the same as that in the study of the high temperature layered 
superconductors by taking the 2D limit of the work of \cite{Pierson-J}
for the case of a vortex gas with a current applied uniformly through all
layers. 
This current can be associated with the applied stress in the
dislocation systems. Comparing the Hamiltonian of the two models one can 
identify the corresponding parameters between them.  
This identification makes possible to obtain the recursion relations for 
dislocation systems by using the recursion relations found 
 in the case of layered superconductors in the 2D limit 
(see Pierson's paper in Ref.\  \cite{Pierson-J}).
Pierson in his work performed a mathematically rigorous, 
real-space renormalization group (RG) analysis following the original 
approach used first by Kosterlitz 
\cite{Kosterlitz}. 

Of course, one can derive the recursion equations independently of 
using the above analogy. Since in Pierson's work the derivation of the
recursion relations are not presented (only the results) we performed  
the RG analysis with our dislocation systems for clarity. We followed the 
procedure used in Ref.\ \cite{Young} (the details to be published 
somewhere else).
 
In this letter we present a consistent treatment of the dislocation
systems in the presence of applied stress. 
We use the same model as in Ref.\ \cite{KhPV-PRL} (Eq.\ (1) therein) and 
give the new recursion relations
which includes the effect of the applied external stress 
both in the fugacity (as in the KPV model) and in the coupling constant.  
In our calculations the stress is taken in lowest nonvanishing order 
both in the equation of the fugacity and the coupling constant.
From this study it turns out that the equation for the fugacity 
remains unchanged (same as the second equation of (4) 
in Ref.\ \cite{KhPV-PRL}), while in the equation for the coupling constant 
an additional term  proportional to the square of the stress appears. 
Therefore, our calculations reveal that the indirect
effect of the stress modifies only the recursion relation for the coupling
constant. However, this latter change has several consequences which will
be discussed below.    

In the KPV model the stress dependence of the transition temperature is
given by a nonlinear implicit equation derived from the stability
analysis of the recursion equations (see Eq.\ (6) in Ref.\ \cite{KhPV-PRL}). 
Using our new recursion relations a different  nonlinear equation is found
for the stress dependence of the transition temperature. The above
mentioned indirect effect of the stress can be associated precisely. 
Then, for different materials, we compare the results obtained from the KPV 
model with that found by using our recursion relations. 
Qualitative explanations are given for the deviations of the transition 
temperature from the KPV model. Finally, we summarize the necessary steps 
for improving the model in order to treat dislocation systems 
more realistically. 

A consistent treatment of the applied stress 
and the role of the indirect effect of the stress in the dislocation 
systems are our main subject in this paper. Our results obtained from the 
extension of the KPV model are a new contribution in this field.

In the frame work of the continuum elasticity theory \cite{Nabarro}
using the same interaction energy between dislocations as in the KPV model 
(Eq.\ (A1) in Ref.\ \cite{KhPV-long-I})
the Hamiltonian of the system ${\cal {H}}$ containing $N$ number of 
dislocations may be expressed as
\begin{eqnarray}
 H =  -\beta{\cal {H}}
  &&=  2\pi K_0\sum_{<i,j>} p_i p_j G\left ({\bf r}_{ij}\right ) 
 +  E_0 \sum_{i}^{N} p_i W({\bf r}_i) \nonumber  \\
 && + \ln y_0 \sum_{i}^{N} p_i^2, 
\label{hamilton}
\end{eqnarray}
where $\beta=1/{k_{\rm B}T}$ and 
${\bf r}_{ij} = {\bf r}_i -{\bf r}_j $ is the distance between 
the $i$th and $j$th dislocation. 
The summation over pairs $<i,j>$ in the Hamiltonian assumes 
$i \ne j $ and counts each pair just once. The charge $p_i=+1$ if the 
Burgers vector at site $i$ is directed to 
the positive x-axis and equals to $-1$ if it is pointed to 
the negative x-axis. The charges satisfy the 
neutrality condition $\sum_i p_i =0$. 

The interactions between dislocations is  given by the first term 
in (\ref{hamilton}) where  
\begin{equation}
G\left ({\bf r}\right ) = 
\ln \left(\left |{\bf r}\right |/a_0 \right)
\end{equation}
and $a_0$ is the cut-off 
radius. The coupling constant in (\ref{hamilton}) 
$K_0  =  \frac{\beta J}{2\pi}$ and 
$J  = \frac{\mu_0 B_0}{\mu_0+B_0}\,\frac{b_0^2}{\pi},$
where $\mu_0$ and $B_0$ 
are the shear and bulk 
moduli in the absence of dislocations and $b_0$ is the magnitude of the 
Burgers vector.

The second term in the Hamiltonian is a sum of the interactions
between dislocations and the applied external stress 
that can be derived from the Peach-Koehler formula \cite{PeachKoehler,Landau}. 
This interaction 
may be written in such a form that is equivalent to the interaction energy 
$p_i{\bf E }{\bf r}_i$ between 
the 'charge' $p_i$ and the homogeneous external 'electric field' 
${\bf E}  =  \sigma\beta b_0{\bf \hat{x}},$ where 
${\bf \hat{x}}$ is the unit vector pointed to the positive $x$-axis. Then  
$E_0$ in the Hamiltonian is related to the applied external stress 
by 
\begin{equation}
E_0= a_0 |{\bf E}|=\sigma\beta b_0 a_0
\end{equation}
 and the interaction potential 
$W$ is given by  
\begin{equation}
W({\bf r}_i) = \frac{1}{a_0}\, {\bf r}_i {\bf \hat{x}} 
- \frac{1}{2}.
\end{equation}
 Note that a constant $-1/2$ has been added to $W$ to be 
consistent with the definition of the core energy $E_c$ of the dislocation 
(which must be one half the energy of a dislocation pair at smallest 
separation and oriented along the direction of the external field 
${\bf E}$). 

Finally, in the last term of the Hamiltonian, the fugacity 
$y_0$  is related to the core 
energy by 
\begin{equation}
\ln y_0 = -\beta E_c + E_0/2,
\end{equation}
where the second term is due to 
the shift of the potential $W$. 

It is easy to show that our Hamiltonian is 
equivalent to that of the KPV model and in this form one can also recognize 
the resemblance to the Coulomb gas model with applied external electric field.
Furthermore, it is clear that the Hamiltonian (\ref{hamilton}) is similar
to that of the model of the layered superconductors if we omit the terms
containing $\lambda$ which relates to the interlayer coupling 
(see Eq.\ (1) -- (3) in Ref.\  \cite{Pierson-J}). Below we make a precise
correspondence between them. 

First, as an extension of the KPV model we derive the recursion equations 
including the effect of the stress both in the equation of the fugacity
and the coupling constant. 

A possible starting point for such a calculation is the partition function.
Following the procedure used in Ref.\ \cite{Young} 
one can derive the recursion relations 
(the details to be published somewhere else). 

However, using the above discussed analogy between our dislocation systems
and the layered superconductors we may obtain our recursion relations by
simple identifying the parameters in the two models via the two
Hamiltonians. We found the following correspondences between the two
models (the parameters in Ref.\ \cite{Pierson-J} are on 
the left hand side, while our parameters are on the right hand side)
\begin{eqnarray}
\tau & \rightarrow & a, \,\,\,\,\, \epsilon \rightarrow  l, 
\,\,\,\,\, \beta p^2  \rightarrow  2\pi K, \\
x & \rightarrow & \frac{2}{\pi K}-1,  
\,\,\,\,\, y  \rightarrow  2\pi y, 
\,\,\,\,\, J\beta \tau p_i  \rightarrow  Ep_i,
\end{eqnarray}
where $J$ does not contain the absorbed factor of $\sqrt{\beta}\tau$
as in Ref.\ \cite{Pierson-J} after Eq.\ (7). Note that $p_i^2=1$ in our case. 

Using the above identifications and the recursion relations given 
in Ref.\ \cite{Pierson-J} and omitting the $\lambda$ terms
we find 
for the coupling constant $K$, the fugacity $y$ and $E$
\begin{eqnarray}
\frac{dK^{-1}}{d l} & = & 
4\pi^3 y^2 \left( 1+ \frac{E^2}{4}\right), 
\label{K-scale} \\
\frac{dy}{d l} & = & \left (2-\pi K+\frac{E}{2}\right )y, 
\label{y-scale} \\
\frac{dE}{d l} & = & E. 
\label{E-scale}
\end{eqnarray}

Eqs.\ (\ref{K-scale})--(\ref{E-scale}) are our final coupled differential 
equations with the initial conditions  $K(l=0)=K_0$, $E(l=0)=E_0$ and 
$y(l=0)=\exp(-\beta E_c + E_0/2)$. 

The indirect effect of the stress in the coupling constant is reflected by
the second term in (\ref{K-scale}). Omitting the $E^2/4$ term 
in Eq.\ (\ref{K-scale}) 
we recover the same recursion equations as in the KPV model.
The equation for the fugacity given in (\ref{y-scale}) is the same as in 
the KPV model. 
Therefore, the inclusion of the indirect effect of the stress 
changes only the renormalization of the coupling constant $K$.
It is also clear that this indirect effect in $K$  appears in second order
in $\sigma$ since $E \sim \sigma$.

At fixed stress the flows obtained from our recursion relations show that 
below at certain temperature $T_c$ the fugacity $y \rightarrow \infty$ 
and for $T> T_c$ $y \rightarrow 0$. The fugacity $y$ is related to the 
mean density of the unbound dislocation pairs in the system. 
Increasing the applied stress (note that $E_0 \sim \sigma$) $T_c$ 
decreases. According to our recursion relations the reason is threefold: 
(i) the stress increases the fugacity $y_0$, (ii) the last term in 
Eq.\ (\ref{y-scale}) has positive feedback for $y$, 
(iii) in Eq.\ (\ref{K-scale}) the coupling constant $K$ lowers, 
i.e., the dislocation pairs are more weakly bounded. This latter effect,
which is our central issue in this letter, is
related to the above mentioned indirect effect of the applied stress.  

In the KPV model a nonlinear equation was derived for the stress 
dependence of the transition temperature by linearizing the recursion 
equations around the fixed point. Similar way, using 
Eqs.\ (\ref{K-scale})--(\ref{E-scale})
we have found that the dimensionless transition temperature 
$t_c=k_{{\rm B}}T_c/J$ is a solution of the following nonlinear equation 
\begin{eqnarray}
\lefteqn{t_c  =  \frac{1- 2\pi C(E_0) 
\exp\left(-\beta E_c+E_0/2 \right)}  
{4+ E_0},   \label{tc-e2}   }\\
&& {\rm where} \, \, \, C(E_0)  =  \sqrt{1+\frac{E_0^2}{4}}, 
\,\, \, \, \, \, E_0=\frac{\sigma^{\prime} b^{\prime}}{t_c}   \nonumber 
\end{eqnarray}
and $\sigma^{\prime}=\sigma\frac{a_0^2}{J}$ is the dimensionless stress while 
$b^{\prime}$ is the magnitude of the Burgers vector in units of $a_0$.
In the KPV model $C(E_0)=1$. 
The indirect effect of the applied stress 
in Eq.\ (\ref{tc-e2}) appears via the  $E_0^2$ term in $C(E_0)$.
The expressions for $t_c$ obtained from the two models differ only by 
the parameter  $C(E_0)$. 

\begin{figure}
{\centerline{\leavevmode \epsfxsize=8cm \epsffile{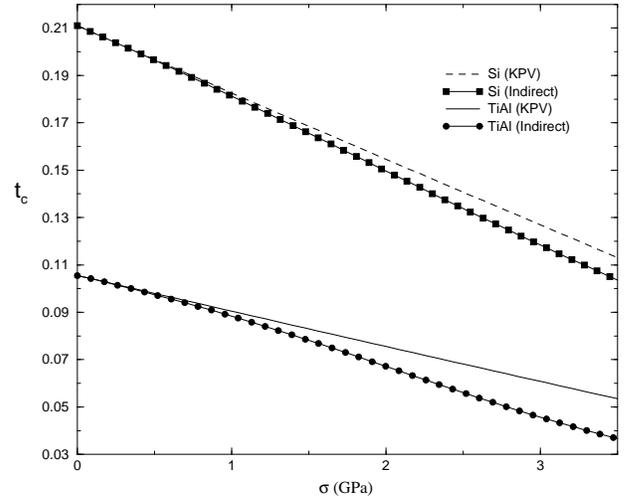}}}
\caption{The transition temperature, $t_c$ as a function of the applied 
stress, $\sigma$ for TiAl and Si. 
For parameters see the text. $t_c$ and the stress are units of 
$J/k_{\rm B}$ and GPa, respectively.
In case of TiAl the solid line and the line with circles correspond to 
the KPV and our model, respectively. In case of Si the dashed line 
and the line with squares are the results from the KPV and our model, 
respectively.
\label{tc-plot}}
\end{figure}

To study the role of the indirect effect of the stress as an 
extension of the KPV model, we solved numerically Eq.\  (\ref{tc-e2}) 
for the transition temperature $t_c$ at different 
values of $\sigma$. We have chosen three particular materials, 
namely, TiAl, Si and NiAl with the same material parameters as 
in \cite{KhPV-long-II}. According to these parameters the core energies 
$E_c=0.25J$ for TiAl, $E_c= 0.78J$ for Si and $E_c=0.27J$ for NiAl 
were used.  
Fig.\ \ref{tc-plot} shows the variation of $t_c$ with the applied stress, 
$\sigma$ obtained from our calculations and from the KPV model 
for TiAl and Si. For comparison, the transition temperature is plotted 
in unit of $J/k_{\rm B}$, while we used the same stress range for 
both materials.
 
The result from the KPV model can be calculated 
by simply taking $C(E_0)=1$ in Eq.\ (\ref{tc-e2}). 
It is important to mention that in both the KPV and our model it is 
assumed that the fugacity, $y_0^2 \ll 1$, i.e., the average density of 
dislocations is small enough. From our calculation it turns out that 
for $\sigma <3.5$ GPa $y_0^2 < 1$.
For all the three materials the fugacity was always lower 
in our model comparing with the KPV model.

From Fig.\ \ref{tc-plot} one can see that the indirect effect of 
the stress results in a small deviation from the original 
KPV model for Si, especially for small values of stress.
Increasing the stress the differences become larger 
but it is still about 7\% at $\sigma = 3$ GPa.
However, the situation for TiAl is changed substantially. Qualitatively, 
two changes can be seen from the figure. First, the transition temperature 
is smaller for TiAl than for Si for all values of stress 
(it is approximately 0.1 smaller at zero stress).
Second, the transition temperature 
calculated from our model departs from the prediction of the KPV model 
at lower values of $\sigma$ for TiAl comparing with  those for Si. 
The differences in $t_c$ between our  and the KPV model are 
substantially larger for TiAl than in the case of Si 
(the difference is about 25 \% at $\sigma = 3$ GPa for TiAl, 
while only 7\% for Si).  
We have also calculated the stress dependence of $t_c$ for NiAl. We found that 
the qualitative feature remains the same as for TiAl and only a slight change 
can be seen in the values of $t_c$ vs.\ $\sigma$ (for zero stress
$t_c=0.112$ for NiAl). 

The first observation can be understood  
from the core energy dependence of $t_c$ at zero stress, 
using  Eq.\  (\ref{tc-e2}). With increasing $E_c$ the 
transition temperature decreases due to the exponential term 
in Eq.\  (\ref{tc-e2}). 
The physical explanation of this behavior is as follows.  
For large $E_c$ the number of thermally activated dislocations is relatively 
small.  This implies that the screening of the dislocation pairs 
by other dislocations is weaker. The coupling strength (normalized by $J$) 
between pairs remains strong. Hence, the unbounding of the pairs, 
which is related to the phase transition, takes place at higher temperature.
The core energies for TiAl and NiAl are approximately the same, while for Si
it is much larger. Thus, $t_c$ will be higher for Si than TiAl and for NiAl 
it is nearly the same as that for TiAl. 

The second observation is less obvious. Again, the analysis of the 
core energy dependence of $t_c$ shows that  at lower $E_c$ 
the indirect effect of the stress becomes more prominent.
The parameter $C(E_0)$ in Eq.\  (\ref{tc-e2}) can be dominant 
besides the suppression of the exponential term. 
On the other hand, with inreasing stress, $t_c$ obtained from our model 
is always lower than that predicted from the KPV model independently 
from the value of  $E_c$. The physical reason is that at higher $\sigma$ 
the so-called 'electric field' makes weaker the bounds between 
the dislocation pairs.  In Eq.\ (\ref{K-scale}) it is reflected by the 
second term. This term is entirely the consequence of the indirect 
effect of the stress. 

In the KPV model the Burgers vector has only two possible orientations
correspondig to  $+1$ and $-1$ charges. However, in a solid there can
be more orientations of the Burgers vector depending on the crystal 
structure. It is desirable
to extend the model in order to take into account the possible Burgers
vector systems in the crystal.   
In the present study (and also in the KPV model) the role of the angular 
dependent force between dislocations has not been considered. 
The quantative behavior of the stress dependence of the transition 
temperature is likely to influence by this force. 
Some preliminary work on these directions have already 
been made \cite{magan}.  

The KPV model has been applied to the brittle-to-ductile transition 
and the deformation of whiskers in Ref.\  \cite{KhPV-long-II}. 
We believe that our model contributes to a better understanding of 
these two phenomena.

In conclusion, we present an extension of the KPV model by 
taking into account the indirect effect of the applied external stress 
on the solid. Realizing the analogy between our dislocation systems and
the layered superconductors we derive the correspondig 
recursion equations. From these equations we obtain a nonlinear 
equations for the transition temperature. The implication of our model is 
discussed in the case of three particular materials.  In generaly, we 
find that the indirect effect of the stress in $T_c$ plays 
crucial role in some materials depending on their core energies.

The author wishes to thank  M.\ Khantha, S.\ W.\ Pierson 
for helpful discussions.
This work was supported in part by the Hungarian Science Foundation OTKA  
(T17609/T025866/T17493) and by the Royal Society of London 
(supported by Foreign and Commonwealth Office).


\begin{thebibliography}{999}
\bibitem{Groma} I. Groma, Phys. Rev. B {\bf 56}, 5807 (1997).
\bibitem{KhPV-PRL} M. Khantha, D. P. Pope and V. Vitek, 
Phys. Rev. Lett. {\bf 73}, 684 (1994).
\bibitem{KhPV-long-I} M. Khantha and V. Vitek, Acta Mater. 
{\bf 45}, 4675 (1997).
\bibitem{KhPV-long-II} M. Khantha P. Pope and V. Vitek, Acta Mater. 
{\bf 45}, 4687 (1997).
\bibitem{KhPV-Scripta-1}  M. Khantha, D. P. Pope and V. Vitek, 
Scr. Metall. Mater. {\bf 31}, 1349 (1994).
\bibitem{KhPV-Engin}  M. Khantha, D. P. Pope and 
V. Vitek, Mat. Sci. and Eng. A, {\bf 192/193}, 435 (1995). 
\bibitem{KT} J. M. Kosterlitz and D. J. Thouless, J. Phys. C {\bf 6}, 
1181 (1973).
\bibitem{Kosterlitz} J. M. Kosterlitz, J. Phys. C {\bf 7}, 1046 (1974).
\bibitem{Berez} V. L. Berezinskii, Zh. Eksp. Teor. Fiz. {\bf 59}, 907 (1970)  
[Sov. Phys. JETP {\bf 32}, 493 (1971)].
\bibitem{NelsonHalperin} B. I. Halperin and D. R. Nelson, Phys. Rev. Lett. 
{\bf 41}, 121 (1978); D. R. Nelson, Phys. Rev. B {\bf 18}, 2318 (1978); 
D. R. Nelson and B. I. Halperin, Phys. Rev. B {\bf 19}, 2457 (1979).
\bibitem{Young} A. P. Young, Phys. Rev. B {\bf 19}, 1855 (1979).
\bibitem{Pierson} S. W. Pierson, Phys. Rev. B {\bf 51}, 6663 (1995).
\bibitem{Pierson-J} S. W. Pierson, Phys. Rev. Lett. {\bf 74}, 2359 (1995). 
\bibitem{Minnhag} P. Minnhagen, Rev. Mod. Phys. {\bf59}, 1001 (1987).
\bibitem{Lubensky} P. M. Chaikin and T. C. Lubensky, 
{\it Principles of Condensed Matter Physics\/} 
(Cambridge University Press, 1995).
\bibitem{Mats} I. Tupitsyn, M. Wallin, A. Rosengren, 
Phys. Rev. B {\bf 53}, R7614 (1996).
\bibitem{Nabarro} F. R. N. Nabarro, {\it Theory of Dislocations} 
(Clarendon, New York, 1967).
\bibitem{PeachKoehler} J. S. Peach and J. M. Koehler, 
Phys. Rev. {\bf 80}, 436 (1950). 
\bibitem{Landau} L. D. Landau and E. M. Lifshitz, 
{\it Theory of Elasticity} (Pergamon, New York, 1970).
\bibitem{magan} J.\ Cserti and A.\ Csord\'as: under preparation. 
\end{thebibliography}
\end{document}